\documentclass[ref]{raa}            
\usepackage{graphicx,times}
\usepackage{booktabs}
\usepackage{natbib}
\usepackage{tabularx}
\usepackage{threeparttable}
\providecommand{\bm}[1]{\mbox{\boldmath$#1$\unboldmath}}

\begin{document}

   \title{A Comparison of Approaches in Fitting Continuum SEDs
}

 \volnopage{ {\bf 2012} Vol.\ {\bf X} No. {\bf XX}, 000--000}
   \setcounter{page}{1}

   \author{Yao Liu
      \inst{1,2,3}
    \and David Madlener
      \inst{3}
   \and Sebastian Wolf
      \inst{3}
    \and Hongchi Wang
      \inst{1}
   }

   \institute{Purple Mountain Observatory, \& Key Laboratory for Radio Astronomy, Chinese Academy of Sciences, 2 West Beijing Road,
              Nanjing 210008, China; {\it yliu@pmo.ac.cn}\\
   \and
   Graduate School of the Chinese Academy of Sciences, Beijing 100080,
   China \\
   \and
   Institut f$\ddot{\rm{u}}$r Theoretische Physik und Astrophysik, Universit$\ddot{\rm{a}}$t zu Kiel, Leibnizstr. 15, 24118 Kiel, Germany\\
\vs \no
   {\small Received [year] [month] [day]; accepted [year] [month] [day] }
}

\abstract{We present a detailed comparison of two approaches, the use of a pre-calculated database
          and simulated annealing (SA), for fitting the continuum spectral energy distribution (SED)
          of astrophysical objects whose appearance is dominated by surrounding dust. While 
          pre-calculated databases are commonly used to model SED data, only few studies to date 
          employed SA due to its unclear accuracy and convergence time for this specific problem.
          From a methodological point of view, different approaches lead to different
          fitting quality, demand on computational resources and calculation time.
          We compare the fitting quality and computational costs of these two approaches 
          for the task of SED fitting to provide a guide to the practitioner to find a compromise between desired 
          accuracy and available resources. To reduce uncertainties inherent to real datasets, we introduce 
          a reference model resembling a typical circumstellar system with 10 free parameters. We derive the
          SED of the reference model with our code \texttt{MC3D} at 78 logarithmically distributed wavelengths 
          in the range $[\rm{0.3\,\mu{m},1.3\,mm}]$ and use this setup to simulate SEDs for the database and SA.
          Our result shows directly the applicability of SA in the field of SED modeling, since the algorithm 
          regularly finds better solutions to the optimization problem than a pre-calculated database. As both 
          methods have advantages and shortcomings, a hybrid approach is preferable. While the database provides 
          an approximate fit and overall probability distributions for all parameters deduced using Bayesian 
          analysis, SA can be used to improve upon the results returned by the model grid.
  }

   \authorrunning{Yao Liu et al. }            
   \titlerunning{A comparison of approaches in fitting continuum SEDs} 
   \maketitle

\section{Introduction}
The continuum spectral energy distribution (SED) is an important observable of astrophysical
sources embedded in a dusty environment such as young stellar objects (YSO), active galactic nuclei (AGN),
and post-AGB stars. It allows one to probe the mass, composition, temperature, and spatial distribution of the 
dust. The common method of analysis is the comparison of available observations with predictions derived
from solving the radiative transfer self-consistently with a model describing the dust properties and its spatial
distribution. The task is to find a parameter set that reproduces the observations best for a given model. 
This minimization of the discrepancy between observation and prediction is an optimization problem and normally 
called a {\it fitting procedure}. 

Various fitting algorithms have been proposed and implemented before (e.g., \citealt{press1992}). From a methodological 
point of view, the fitting approaches differ in the quality of the resulting fit and demand on computational resources. 
The most common method is based on a pre-calculated model database that is established on a huge grid in a 
high-dimensional parameter space (e.g., \citealt{robitaille2006, woitke2010}). Once the database is 
established, the optimum parameter set is readily identified by evaluating the merit function, i.e. for our purpose 
the $\chi^{2}$-distribution. However, as the number of grid points increases substantially with the dimensionality of 
the parameter space, the model grid is always a compromise between finite computational resources and 
resolution. {\it Simulated Annealing} (SA) is a versatile optimization 
technique based on the Metropolis-Hastings algorithm that can be used to search for the
optimum of a merit function in arbitrary dimensions (e.g., \citealt{kirkpatrick1983, madlener2012}). 
The main idea is to construct a random walk through parameter space thereby improving the agreement between 
observation and prediction gradually by following the local topology of the merit function. A drawback of this 
method is that no upper bound for the step count to reach the global optimum can be given. Moreover, the 
sequential execution of the algorithm and possible slow convergence of the Markov chain can make SA time consuming. 

In the context of SED modeling, a pre-calculated database is quite commonly invoked to perform the task
because it can not only provide an approximate fit but also enables to evaluate the overall 
uncertainty of all parameters by Bayesian analysis (\citealt{lay1997}). 
On the contrary, only a small sample of studies to date make use of SA in this area due to its unclear accuracy 
and typical convergence time. A main limitation is the significant computing time per individual 
SED model, especially when simulating the SEDs in an optically thick system.
With the recent advance in computing performance, the SA approach is now applicable. The motivation 
behind this study is to evaluate advantages and shortcomings of these two methods when applied to an 
astrophysical optimization problem. We set up an unbiased benchmark by deriving synthetic observations 
from a known reference model to exclude any influence due to model uncertainties on the 
optimization process. We will present fitting quality and computational cost for this 
idealized optimization task using both approaches to provide a guide for practitioners 
to find a compromise between desired accuracy and available resources.

\section{Reference Model and Modeling}
\label{sect:toy_modeling}Circumstellar disks surrounding YSOs, are
considered as an essential step of the star-forming process.
A lot of attention have been paid to these interesting objects,
since they are most probably the birthplace of planet systems
(e.g., \citealt{mundy2000, meyer2007}). The planet formation
mechanism in these disks and the properties of the resulting planetary system 
depend on the structure of the protoplanetary disks that can be constrained by SED modeling. 
In this section we introduce a reference model (RM) to mimic a virtual object 
located in the Taurus star formation region at a distance of $140\,\rm{pc}$ 
and describe our simulation technique. By using a virtual object, all uncertainties 
in regard to the model itself are eliminated.

\subsection{Disk Structure}
We employ a parametrized flared disk in which dust and gas are well mixed and homogeneous throughout the system.
This model has been successfully used to explain multi-wavelength observations of protoplanetary disks like the 
SED and high resolution images (e.g., \citealt{wolf2003butterflystar, schegerer2008, sauter2009}).
For the dust in the circumstellar disk we assume a density structure with a Gaussian vertical profile: \\
\begin{equation}
\rho_{\rm{dust}} \sim R^{-\alpha}\exp\left[-\frac{z^2}{2h^2}\right] \\
\label{dust_density}
\end{equation}
and a power-law distribution for the surface density
\begin{equation}
\Sigma(R) \sim R^{-p},
\end{equation}
where $R$ is the distance from the central star measured in the disk midplane. The proportionality 
factor is determined by normalising the total dust mass in the disk. The disk scale height $h(R)$ follows 
the power law
\begin{equation}
h(R) = h_{100}\left(\frac{R}{100\rm{AU}}\right)^\beta,\\
\end{equation}
with the flaring exponent $\beta$ describing the extent of flaring and the scale height $h_{100}$ at a distance 
of $100\,\rm{AU}$ from the central star. 

Table \ref{tab:referencemodel} lists the parameters of the RM. We truncate the disk at $300\,\rm AU$, a typical size 
found for T Tauri disks and fix the value of $h_{100}$ to $10\,\rm AU$ (e.g., \citealt{andrews2007a}). 
We consider a total dust mass of $5\cdot10^{-5}\,\rm{\rm{M_{\odot}}}$, corresponding to the typical value found in T Tauri 
disks (e.g., \citealt{beckwith1990, andrews2007b, andrews2009}). We make a standard assumption for the dust-to-gas mass 
ratio, i.e., $m_{\rm{dust}}/m_{\rm{gas}}=1/100$. For the flaring exponent, we adopt the value of 1.25 (e.g., \citealt{dalessio1999}). 
The exponent of the dust density profile $\alpha=3(\beta-\frac{1}{2})=2.25$ is derived from 
viscous accretion theory (\citealt{shakura1973}).

\begin{table}[h]
\caption{Parameters of the reference model (RM) and the best-fit in our pre-calculated database.}
\label{tab1}
\centering
\begin{tabular}{l c c}
\toprule
\toprule
 parameter       &   RM  &   best-fit model   \\
  \hline
 $T_{\star}[K]$  & 4000  &  4262     \\ 
 $L_{\star}[\rm{L_{\odot}}]$ & 0.92 & 2.7  \\
 $R_{\rm in}$[AU]   & 2.0   & 2.9  \\
 $R_{\rm out}$[AU]  & 300  & 450  \\
 $m_{\rm{dust}}[\rm{M_{\odot}}]$  & $5\cdot10^{-5}$  & $3\cdot10^{-5}$  \\
 $\alpha$  &  2.25  & 1.61  \\
 $\beta$   &  1.25  & 1.033   \\
 $h_{100}[\rm AU]$ & 10 & 10.5  \\
 $a_{\rm max}[\mu{\rm m}]$  & 2.5  &  25  \\
 $i[^\circ]$ & 60  & 75  \\
\toprule
\end{tabular}
\label{tab:referencemodel}
\end{table}

\subsection{Stellar Heating}
There are several heating sources of the circumstellar disks, such
as irradiation by the central star, disk accretion and turbulent
processes within the disks. To keep our model simple and
decrease the number of free parameters, we consider a passive disk
in which only stellar irradiation is taken into account (e.g., \citealt{chiang1997}). 
We assume parameters of a typical T Tauri star for the central source: $R_{\star}=2\,R_{\odot}$ and
$T_{\star}=4000\,\rm K$, corresponding to a bolometric luminosity of
${\sim}0.92\,L_{\odot}$ (e.g., \citealt{gullbring1998}).

\subsection{Dust Properties}
We consider the dust grains to be homogeneous spheres, since the assumption of spherical grain
is a valid approximation to describe the scattering behavior compared with a more complex and 
fractal grain structure. The dust grain ensemble incorporates both silicate and
graphite material with relative abundances of 62.5\% astronomical silicate and 37.5\% graphite.
To calculate the optical properties of the dust with the Mie scattering-theory, we use the complex refractive
indices of ``smoothed astronomical silicate'' and graphite published by \cite{weingartner2001}.
For graphite, we adopt the common ``$\frac{1}{3}:\frac{2}{3}$'' approximation (\citealt{draine1993}), 
which means the extinction efficiency factor is computed by:
\begin{equation}
Q_{\rm{ext,graphite}}=\frac{1}{3}Q_{\rm{ext}}(\epsilon_{\parallel})+\frac{2}{3}Q_{\rm{ext}}(\epsilon_{\perp}),
\end{equation}
where $\epsilon_{\parallel}$ and $\epsilon_{\perp}$ are the components of the graphite dielectric tensor for the
electric field parallel and orthogonal to the crystallographic axis, respectively.

We assume a power law grain size distribution
$n(a)\propto{a^{-3.5}}$ with
$a_{\rm{min}}\leq{\rm{a}}\leq{a_{\rm{max}}}$, where $a$ represents
the grain radius and $a_{\rm{min}}$ and $a_{\rm{max}}$ are the minimum and the maximum grain radii. 
The size distribution with $a_{\rm{min}}=5\,\rm{nm}$ and $a_{\rm{max}}=0.25\,\mu{\rm{m}}$ is the 
well-known MRN distribution found for the ISM (\citealt{mathis1977}). For the RM, we 
keep $a_{\rm{min}}=5\,\rm{nm}$ and increase the maximum grain size to $a_{\rm{max}}=2.5\,\mu{\rm m}$ 
to account for dust growth in circumstellar disks (e.g., \citealt{sauter2009, ricci2010}).

\subsection{Radiative Transfer Simulation Code}
To derive the observables of the RM, we use the well-tested radiative transfer code MC3D developed by
\cite{wolf2003mc3d}. Based on the Monte-Carlo method, MC3D solves the radiative transfer problem 
self-consistently. It implements the immediate temperature correction technique as described by 
\cite{bjorkman2001} and the continuous absorption concept as introduced by \cite{lucy1999}.
Multiple and anisotropic scattering is considered in the simulations.

\begin{figure}[!htp]
\centering
\includegraphics[width=\textwidth]{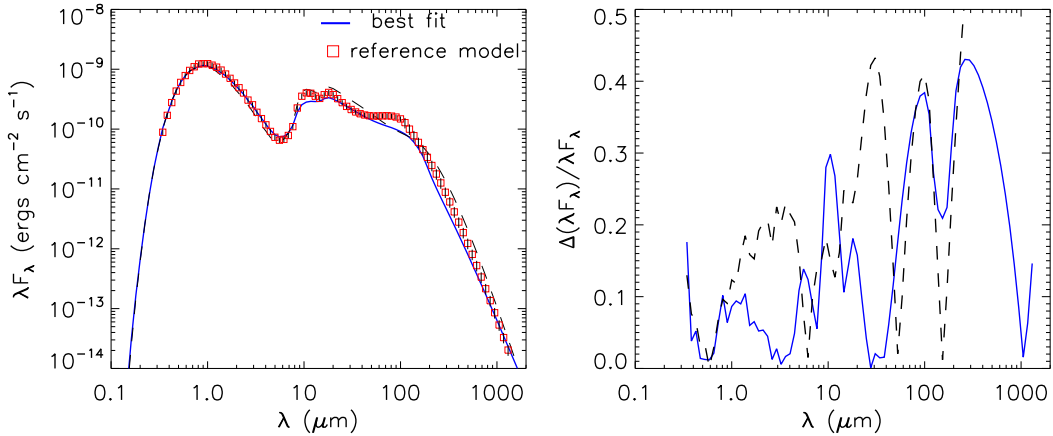}
\caption{{\it Left plot}: the top two model fits in the database. The best-fit model is indicated as a 
blue solid line and the red squares represent the SED of the RM. {\it Right plot}: the moduli of flux 
discrepancies between the models and the RM.}
\label{fig:bestfitofdatabase}
\end{figure}

\subsection{The SED of the RM}
The radiative transfer problem is solved at 100 wavelengths, logarithmically distributed in the 
wavelength range $[\rm 50\,nm,2.0\,mm]$. The red squares in Fig. 1 show the simulated SED of 
the RM at 78 wavelengths in the range $[0.34\,\mu{\rm{m}},1.3\,\rm{mm}]$. Since data points 
within this range can be obtained by current telescopes, we therefore only reproduce fluxes 
at these wavelengths in the fitting procedure.

\section{SED-FITTING WITH A PRE-CALCULATED DATABASE}
A popular approach to analyze SEDs is to solve the radiation transfer
equation based on dust properties and a density distribution. Given a particular
model and radiative transfer code, a database of model SEDs can be established on 
a part of the model's parameter space (\citealt{woitke2010}).

The main advantage of this approach is that it allows to explore
how specific data points influence parameters of the fit. As an example, given a particular dust model 
one (sub)millimeter data point can constrain the total dust mass. An additional data point can help to constrain 
the maximum dust radius in protoplanetary disks. Moreover, observations of a large 
sample of objects can be fitted very fast to a pre-calculated database.

In order to examine the fitting accuracy of a pre-calculated database, we
adapted the MC3D code to the GPU cluster at Purple Mountain Observatory and built a database 
of 70,000 model SEDs using the same dust density profile and grain composition as described in 
Section 2. We used a similar strategy as implemented by \cite{robitaille2006} to sample 
the parameter values in wide space to avoid giving any attention to a particular model, especially the RM. 
Generally, the model SEDs span a reasonable range of parameters constrained by theory and observations 
of protoplanetary disks. The effective temperature $T_{\star}$ and radius of the central star were 
derived by interpolating pre-main-sequence evolutionary tracks given by \cite{siess2000}
with the mass and age randomly sampled and logarithmically spaced in the ranges of $[0.5\,\rm{M_{\odot}},1.5\,\rm{M_{\odot}}]$
and $[0.1\,\rm{Myr},20\,\rm{Myr}]$ respectively. The dust mass was sampled from a wide range of $[10^{-9},10^{-3}]\rm{M_{\odot}}$,
but values between $10^{-7}\,\rm{M_{\odot}}$ and $10^{-4}\,\rm{M_{\odot}}$ were randomly selected more frequently.
The model grid contains a majority of samples with disk outer radius between $100\,\rm{AU}$ and $400\,\rm{AU}$. 
Other reasonable values consistent with observations are also considered. For one half of
our models, we set the disk inner radius to the dust sublimation radius $R_{\rm{sub}}=R_{\star}(T_{\rm{sub}}/T_{\star})^{-2.085}$, 
where we take $T_{\rm{sub}}=1500\,\rm{K}$ to be the dust sublimation temperature (\citealt{whitney2004}). 
The flaring exponent for these models is set to a smaller value than that found for T Tauri
disks, i.e., $\beta<1.25$. This was done in order to interpret the observations of homologously depleted 
disks in which material dissipation is thought to occur throughout the disk 
simultaneously (e.g., \citealt{currie2009}). In the remaining half of models, the inner radius of 
the disk was increased to account for canonical transition disks that are expected to have large inner 
holes due to the mechanism of inside-out disk dissipation (e.g., \citealt{muzerolle2010}). 
The flaring exponent $\beta$ for these models are uniformly sampled between 1.10 and 1.30. 
The exponent of the density distribution $\alpha$ is calculated using the relationship $\alpha=3(\beta-\frac{1}{2})$ 
derived for a steady-state disk in which the temperature and surface density profile are coupled by $p+q=3/2$,
where $q$ is the exponent of the temperature profile $T{\sim}r^{-q}$.
The models with scale height parameter $h_{100}$ between $8\,\rm{AU}$ and $15\,\rm{AU}$ occupy 
a large portion, and other reasonable values outside this range are also contained in our database.
The entire model grid was established in approximately 30 days using 32 CPUs on the computer cluster.

The RM and the grid in parameter space of the database are set up independently. 
We fit the SED of the RM to this database and examine whether the
database returns a model with parameter values close to the reference. Additionally, the
distance is fixed at $140\,\rm pc$ and no extinction is used to avoid any uncertainties from the employed 
extinction law. In short, we are only interested in the fitting quality of this approach.
Fig. \ref{fig:bestfitofdatabase} shows the top two SED models in the database and the flux 
discrepancies between these models and the RM as well. The blue solid line represents the 
best-fit with a minimum $\chi^{2}=179.5$ calculated from
\begin{equation}
\chi^{2}=\sum_{i=1}^{n}\frac{[F_{i}-F_{i}(\rm{RM})]^{2}}{\Delta{F_{i}^{2}(\rm{RM})}}.
\end{equation}

Here, $n$ is the number of data points, $F_{i}$ is the simulated
flux, $F_{i}(\rm{RM})$ is the ``observed flux''. The photometric
uncertainty, $\Delta{F_{i}(\rm{RM})}$, was assumed to be comparable to typical
observation errors of real instruments working in different
wavelength regimes, $10\%$ and $15\%$ of the flux at shorter and longer wavelengths respectively. 
This assumption has little impact on our study as along as the observation errors are 
not changed. Obviously, the best-fit SED is in good agreement with the SED of the RM. The 
corresponding parameter values of the best model are listed in Table \ref{tab:referencemodel}. It is 
particularly noteworthy that the fitting results constrain the inner radius and dust mass 
quite well. The central star of the best-fit model is much brighter than the RM. However, the fluxes 
at short wavelengths are close to the RM. This can be explained by higher disk 
inclinations (e.g., the best fit $i=75^{\circ}$) for which the outer disk occults the central 
star and even the inner disk regions, leading to a significant extinction for short-wavelength 
emission (\citealt{chiang1999}). The slight flux deficit in the far-infrared domain (${\sim}100\,\mu{\rm m}$) 
can be explained by a smaller flaring exponent ($\beta=1.033$) of the best-fit model
as compared to the RM. Disks with larger flaring exponent intercept a larger portion of the 
central star's radiation due to a larger flaring angle and consequently re-emit more infrared flux. 
The maximum dust grain radius $a_{\rm{max}}=25\,\mu{\rm m}$ is one order of magnitude larger than 
the reference value. This shallows the spectral index of the best-fit model at (sub)millimeter wavelengths 
and weakens the $10\,\mu{\rm m}$ silicate band. However, the database contains only four values for $a_{\rm max}$, 
namely $0.25\,\mu{\rm{m}}$, $2.5\,\mu{\rm{m}}$, $25\,\mu{\rm{m}}$, and $250\,\mu{\rm{m}}$, hence the difference 
in $a_{\rm max}$ between the best-fit model and the RM is only one grid interval.

\begin{figure}[!htp]
\centering
\includegraphics[width=\textwidth]{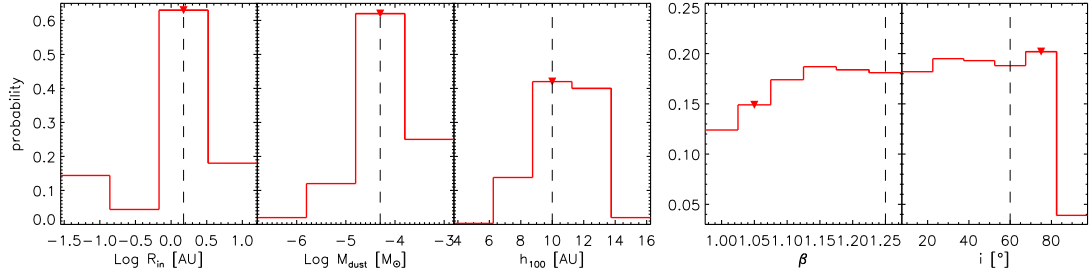}
\caption{Bayesian probability distribution of selected disk parameters. The triangles mark the probability bins
            containing the best-fit parameters. The vertical dashed lines symbolize the reference values.}
\label{fig:bayesian}
\end{figure}

In order to estimate the confidence range for each parameter, we performed a Bayesian analysis 
using the reduced $\chi^2$ defined by
\begin{equation}
\chi_{\rm{red}}^2=\frac{\chi^2}{N_{\rm{data}}-n-1},
\end{equation}
where $N_{\rm{data}}=78$ and $n=10$ are the number of data points and degrees of freedom 
respectively (e.g., \citealt{pinte2007, pinte2008, bouy2008}). The relative probability 
$\exp\left[-\chi_{\rm{red}}^2/2\right]$ of each parameter set is used as a statistical weight to 
calculate the relative likelihood for every parameter by summing over all models with a common 
value and normalising with the total sum. Non-uniform distributions of models in parameter space 
due to different sampling density are considered as the priori probabilities and have to be taken 
into account. Quantitative error bars for every parameter can then be deduced from the 
resulting probability distributions.

Fig. \ref{fig:bayesian} presents the probability distribution for some selected disk parameters. 
The results show that SED analysis places constraints on the parameters differently, e.g. only 
high inclinations (edge-on) can be excluded, while all other configurations appear to be equally 
likely. The scale height $h_{100}$ and the flaring exponent $\beta$, characterising the vertical 
geometry of the disk and hence its capacity to absorb stellar energy, are constrained to some extent. 
Our Bayesian analysis clearly supports the notion that SEDs mainly contain information on the inner radius 
and dust mass of circumstellar disks as we get the best constraints on these two parameters.
Moreover, we noticed that the best-fit model implies a parameter value (e.g., $\beta$) 
offset from the peak of its probability distribution. This is an indicator of model degeneracy 
often encountered in pure SED fitting.

\section{Optimization with Simulated Annealing}
Simulated annealing (SA) is a variation of the Metropolis-Hastings algorithm that applies
methods of statistical physics to optimization problems like the traveling salesman problem and has 
been used for many optimization challenges (e.g., \citealt{metropolis1953, hastings1970, kirkpatrick1983}). 
The physical analogy of quenching a hot melt has given this extremely useful method its name
as the main feature of SA is the gradual reduction in temperature during optimization. Basically, the 
distribution under scrutiny can be interpreted as the inner energy of a system in 
a thermal bath. By starting at a high temperature and subsequently cooling the bath while 
allowing the system to evolve through phase space the system trajectory intrinsically leads to regions of lower 
inner energy and thus to the global optimum of the underlying problem. One of the major advantages of this Monte 
Carlo method is the independence from the dimensionality. 
Local extrema are overcome intrinsically, no gradients have to be evaluated 
and the parameter space can be discrete or continuous. We adapted this technique to search the optimal 
disk model for the synthetic SED derived from the RM.

\begin{table}[!htpb]
 \centering
   \caption{Initial Step Widths and Boundary of the Parameter Space}
   \tabcolsep 4mm
   \begin{tabular}{lccc}
     \toprule
     \toprule
parameter                   &  $\beta_{k}$           &   min        &   max       \\
   \midrule
$T_{\star}$[K]              & $0.2\,T_{\star0}$      &  2500        & 7000        \\
$L_{\star}$[$L_{\odot}$]    & $0.2\,L_{\star0}$      &  0.2         & 8.0         \\
$R_{\rm IN}$[AU]            & $0.2\,R_{\rm{in0}}$    &  0.1         & 20.0        \\ 
$R_{\rm out}$[AU]           & $0.2\,R_{\rm{out0}}$   &  50          & 1000        \\
$\alpha$                    & 0.005                  &  1.5         & 3.0         \\
$\beta$                     & 0.005                  &  1.0         & 1.3         \\
$h_{\rm 100AU}$[AU]         & 0.25                   &  5.0         & 20.0        \\
$m_{\rm dust}$[$M_{\odot}$] & $0.5\,M_{\rm{dust0}}$  & $10^{-9}$    &  $10^{-3}$  \\
$a_{\rm max}$[$\mu{\rm m}$] & $0.2\,a_{\rm{max0}}$   &  0.25        & 1000        \\
$i$[$^{\circ}$]             &  2.0                   &  15          & 85          \\
   \toprule
   \end{tabular}
   \tablecomments{0.6\textwidth}{$T_{\star0}$, $L_{\star0}$, $R_{\rm{in0}}$, $R_{\rm{out0}}$, $M_{\rm{dust0}}$ and $a_{\rm{max0}}$ are the starting points.}
\label{tab:stepsize_bound}
\end{table}

\begin{table}[!htpb]
 \centering
   \caption{Starting Points of the Markov Chains}
   \tabcolsep 3.0mm
   \begin{tabular}{lcccccccc}
     \toprule
     \toprule
parameter                          &   I     &   II   &  III    &  IV   &  V     &   VI   &   VII   &  VIII  \\
   \midrule
$T_{\star}$[K]                     &  4372   &  4040  & 6540    & 2984  & 3277   & 3781   &  3052   &  2806  \\
$L_{\star}$[$L_{\odot}$]           &  0.93   &  0.71  & 5.48    & 0.29  & 0.38   & 0.57   &  0.32   &  0.25  \\
$R_{\rm IN}$[AU]                   &  0.906  &  0.613 & 11.64   & 0.17  & 0.24   & 0.45   &  0.19   &  0.14  \\ 
$R_{\rm out}$[AU]                  &  173.8  &  139.4 & 736.5   & 69.0  & 83.9   & 117.3  &  72.2   &  61.3  \\
$\alpha$                           &  2.124  &  2.013 & 2.846   & 1.661 & 1.759  & 1.927  &  1.684  &  1.602 \\
$\beta$                            &  1.124  &  1.102 & 1.269   & 1.032 & 1.051  & 1.085  &  1.036  &  1.021 \\
$h_{\rm 100AU}$[AU]                &  11.24  &  10.13 & 18.46   & 6.61  & 7.59   & 9.27   &  6.84   &  6.02  \\
$m_{\rm dust}$[$10^{-5}M_{\odot}$] &  0.32   &  0.88  & 0.00041 & 22.0  & 9.2    & 1.9    &  18     &  39    \\
$a_{\rm max}$[$\mu{\rm m}$]        &  7.8    &  4.3   & 428.8   & 0.6   & 1.0    & 2.6    &  0.69   &  0.43  \\
$i$[$^{\circ}$]                    &  44     &  38    & 77      & 22    & 27     & 35     &  24     &  19    \\
   \toprule
   \end{tabular}
\label{tab:startingpoint}
\end{table}

\subsection{Algorithm}
SA creates a Markov chain of points in parameter space by starting from an 
arbitrarily chosen set of parameters within the constrained range, see Table \ref{tab:stepsize_bound}. 
The parameter intervals considered here are consistent with those for establishing the pre-calculated database.
The Markov chain progresses from the starting point by generating new parameter sets depending on the current 
position until ideally the global optimum is reached. The parameter values of each step are 
recorded and can be used for subsequent analysis. We list 8 starting models in Table \ref{tab:startingpoint} with 
parameter values that deviate from the RM in varying degrees. After sampling a uniformly distributed number 
in the range [0, 1] using the \texttt{randomu} function in IDL programming language, we scale it to the corresponding 
parameter range to obtain the starting point in every dimension. The argument \texttt{seed} of the \texttt{randomu} 
function used to initialize the random sequence is chosen to be identical to the chain number. 
Moreover, the random sampling procedure is implemented in logarithmical space for parameters with large dynamical range, such as the 
inner radius, dust mass and maximum grain size. We define the starting points in this way to ensure independence from the RM and reproducible results.
The applicability of SA to the optimization of SEDs of circumstellar disks can be validated by comparing the disk models 
generated by the algorithm with the RM. Moreover, we are interested in the number of Markov chains and steps 
per chain needed to find a good fit. The displacement $\Delta\mathbf{a}$ in parameter space is sampled 
from a gaussian distribution using varying step sizes $\beta_k$:
\begin{equation}
{\rm Prob}(\Delta\mathbf{a}) \sim \prod_k\exp\left[-\frac{\Delta a_k^2}{2\beta_k^2}\right],
\end{equation}
where the index $k\,\in\,\left\{1,...,D\right\}$ enumerates the parameter dimensions.
To ensure randomness of the sampling processes in our study, 4 different seeds were used to initialize the 
pseudo-random number sequence for each of the 8 starting points, resulting in a total number 
of $8\cdot4=32$ different Markov chains.

After sampling a new step $\Delta\mathbf{a}$, we accept or reject it by evaluating the acceptance probability 
using the difference in $\chi^2$ between the last accepted step and the new position:
\begin{equation}
A = \min\left\{1,\exp\left[-\frac{\Delta\chi^{2}}{\tau}\right]\right\}
\label{eqn:accept}
\end{equation}
We immediately accept a new step if the $\chi^{2}$ at the new position is lower than at the actual position. If we
only accept choices with $A=1$, the Markov chain converges to the next local minimum that is not necessarily
identical with the global optimum. Instead, a uniformly distributed number $u\in[0,1]$ is sampled and compared with $A$. 
The new proposed position is accepted if $u<A$ and the chain has the chance to escape a local minimum depending on 
the actual temperature. The crucial role of the temperature $\tau$ is discussed in the next section.

\subsection{Annealing Schedule}
The random walk through parameter space not only moves to better models with smaller $\chi^2$ but also to worse 
models with a probability ${\rm{\sim}\,exp[-\Delta\chi^2/\tau]}$. The system temperature 
$\tau$ is hence the major control parameter of the random walk and is gradually reduced to steer the Markov 
chain to the vicinity of the global optimum. It can be shown in the case of logarithmic cooling that the Markov 
chain will reach the global optimum asymptotically (\citealt{geman1984}). Unfortunately, no upper bound for 
the time to reach the optimum with this strategy can be given and a variety of cooling schedules have been 
proposed to achieve faster convergence (\citealt{nourani1998}). Monotonic cooling, like in exponential and 
linear schedules, is used in most applications of SA. The chain starts at a high temperature, runs 
through a {\it melting phase} to reduce any bias from the chosen starting point, and is then 
successively cooled to reduce the probability of moving to worse parameter sets until it freezes at 
a location in parameter space.

However, monotonic cooling schedules always contain several free parameters that must be adjusted 
empirically to control the duration of the optimization . For this study we implemented a 
non-monotonic schedule by setting the chain temperature $\tau$ after each accepted step to
\begin{equation}
\tau=\gamma\cdot\chi_{\rm{last}}^{2}.
\label{eqn:schedule}
\end{equation}
Here, $\gamma\sim0.25$ is a fixed parameter controling the probability of taking an uphill step 
(e.g., \citealt{locatelli2000}). Without any user interaction during the optimization procedure, 
our approach enables a chain to escape a local minimum. 

\subsection{Adaptive Step Sizes:}
The step size $\beta_{k}$ controls the displacement between the two adjacent positions in parameter space. 
Optimal step sizes cannot be defined at the beginning due to the lack of any information on the distribution 
under investigation. Fixed step width is not an appropriate solution because this approach can waste 
computation time if the chosen value are too large compared to the local shape of the merit function, 
leading to a low efficiency of the algorithm. If the increments are too small, the Markov chain converges 
slowly and the escape from a local valley becomes unlikely as too many steps are needed.
The step sizes for each of the parameters hence have to be adapted along the run. 
For this purpose, the reasonable initial values that are dependent on the starting points of the Markov chain 
are listed in Table \ref{tab:stepsize_bound}.

We define a local acceptance ratio  of the chain using the sequence $\eta_{n}\,\in\,\left\{0,1\right\}$ of
rejected and accepted steps by calculating
\begin{equation}
\xi_{n}=\frac{1}{l}\sum_{m=n-l+1}^n\,\eta_{m}
\end{equation}
for the constant lookback length $l\,<\,n$, where $n$ refers to the step count.
In order to maintain the optimal acceptance ratio in the vicinity of $\xi_{0}=0.234$ (\citealt{roberts1997}), we implement
a control loop that adjusts the step widths $\beta_{k}$ by analysing the last $l=25$ positions
of the chain. The algorithm calculates the moduli of relative changes between the proposed parameter $a_{n,k}$
and the previously accepted parameter $\overline{a}_{n-1,k}$:
\begin{equation}
c_{n,k}=\left|\frac{a_{n,k}-\overline{a}_{n-1,k}}{\overline{a}_{n-1,k}}\right|.
\end{equation}
By separately summing up the relative changes $c_{n,k}$ for every parameter in accepted and rejected proposals,
two vectors $g_{n,k}$ and $b_{n,k}$ can be derived to describe the impact of good and bad decisions:
\begin{equation}
g_{n,k}=\sum_{m=n-l+1}^n\,\eta_{m}c_{m,k},\,\,\,\, b_{n,k}=\sum_{m=n-l+1}^n\,(1-\eta_{m})c_{m,k}
\end{equation}
if $\xi_{n}>\xi_{0}$, the step size of the parameter with the smallest component in $g_{n,k}$ is increased to
encourage riskier proposals. On the contrary, if $\xi_{n}<\xi_{0}$ the step size of the parameter with the largest
component in $b_{n,k}$ is decreased to induce more conservative proposals. In either case the step size of the
selected parameter is adjusted by multiplying with the factor of $(1-\xi_{0})+\xi_{n}$. 
Moreover, for parameters spanning several orders of magnitude, like for $m_{\rm{dust}}$ and $a_{\rm{max}}$, we 
adapt their step widths using a logarithmical scale.

Additionally, the step sizes are controlled to be $<50\%$ and $>1\%$ of the current parameter values in order
to avoid large fluctuations and extremely slow convergence of the Markov chain respectively.

\subsection{Chain Abortion}
Since no upper bound for the step count to reach the global optimum in SA can be given, we must define a 
reasonable criterion to stop an optimization run. When the system temperature $\tau$ drops, the Markov chain can 
get trapped at a location in parameter space if surrounded by sufficiently steep gradients as any deviation from 
the local minimum will yield large $\Delta\chi^2$, see Eq. 8.

\begin{figure}[!htp]
\centering
\begin{minipage}[c]{0.5\textwidth}
\centering
\includegraphics[width=\textwidth]{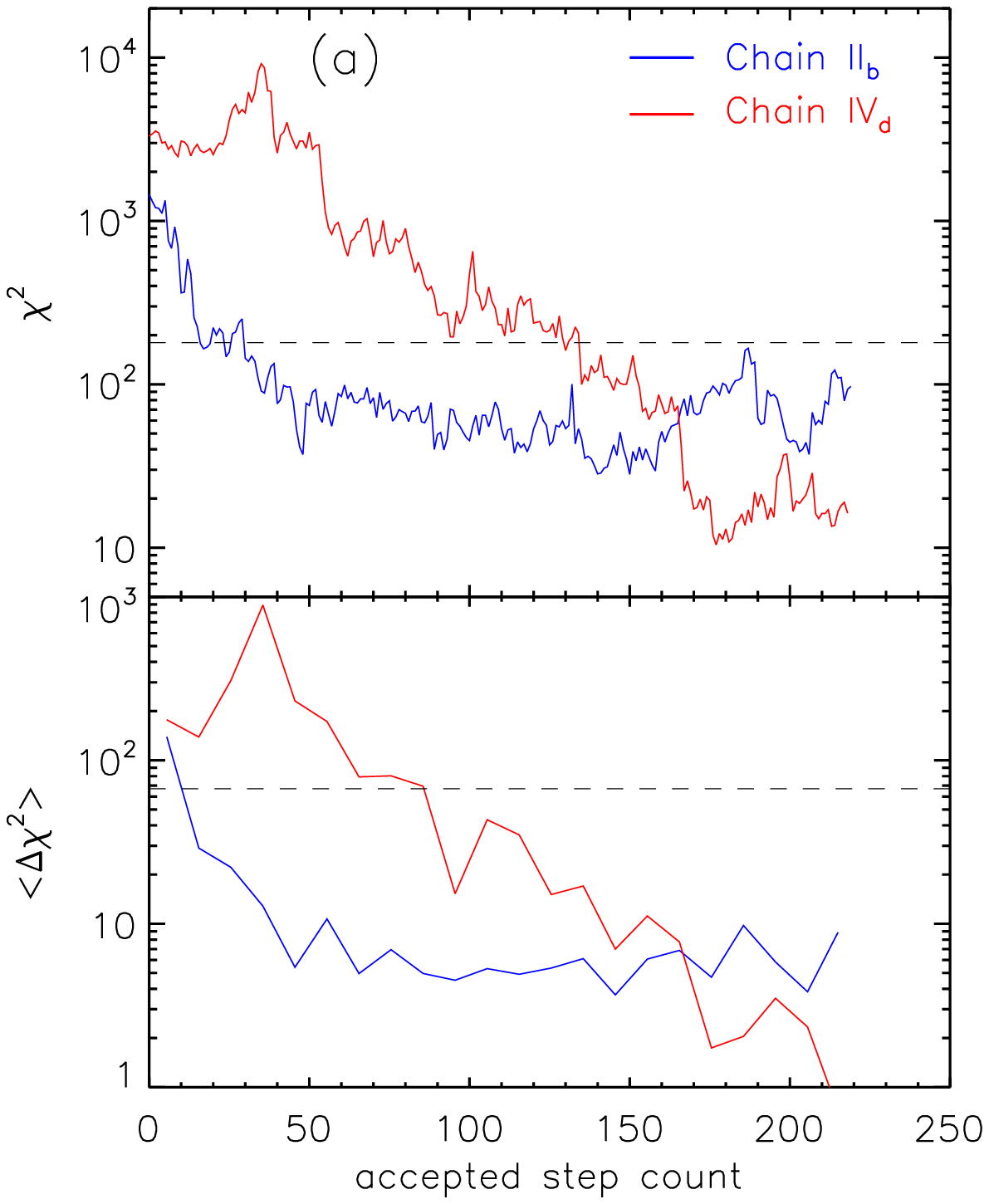}
\end{minipage}
\hfill\hspace*{-1mm}\begin{minipage}[c]{0.5\textwidth}
\centering
\includegraphics[width=\textwidth]{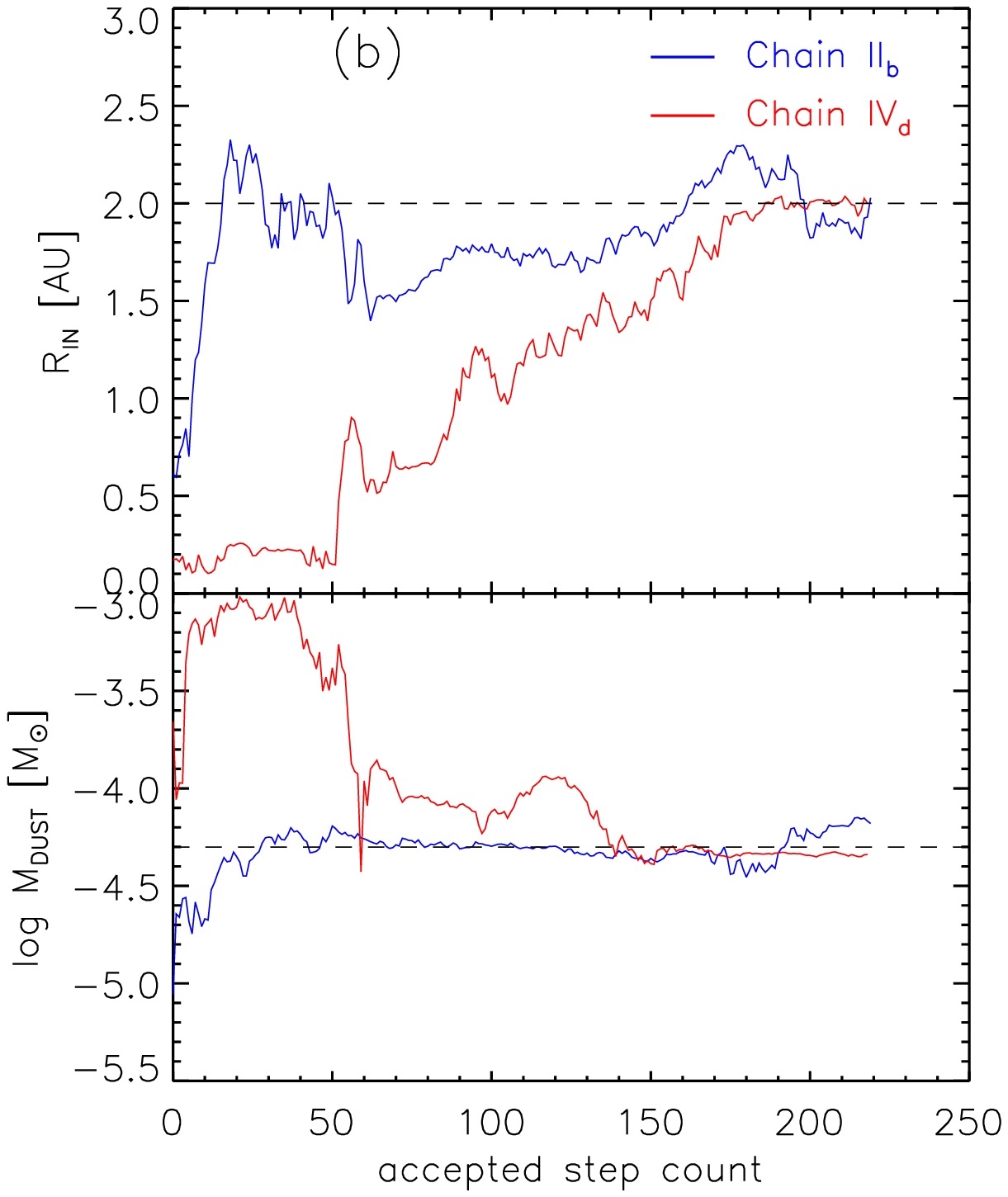}
\end{minipage}
\caption{$\chi^2$ and selected model parameters plotted against the number of accepted steps for the Markov
chains $\rm II_{d}$ and $\rm{IV_{d}}$. The horizontal dash lines in the upper and lower panel of 
diagram (a) depict the best-fit $\chi^{2}$ from the database and the level of $\langle\Delta\chi^2\rangle=67$ 
respectively. Diagram (b) depicts the inner radius (upper panel) and the dust mass (lower panel) of the accepted 
models. The dashed lines in diagram (b) symbolize the reference values.}
\label{fig:tracechain}
\end{figure}

As the system temperature is coupled to the $\chi^2$ of the current position, see Eq. 9. 
Hence, the probability to escape a local minimum does not converge monotonically to zero as the probability to 
climb uphill remains constant. Hence, it is not straightforward to decide when to abort the chain using a 
temperature threshold as normally implemented (e.g., \citealt{nourani1998}). Instead, we analyze the quality 
of the fit step by step for every accepted model of an individual Markov chain. The variation of the fitting 
quality is evaluated by calculating the modulus of the difference $\Delta\chi^2$ between two adjacent accepted steps. 
A particular Markov chain is aborted if the box average $\langle\Delta\chi^2\rangle$ remained below a variation 
threshold for a certain number of steps $N_{\rm{abort}}$ in a row. We used $\Delta\chi^2_{\rm{abort}}=67$ and 
$N{\sim}100$ steps and averaged with a box of 10 samples, see Fig. \ref{fig:tracechain} (a) for examples of 
chain $\rm{II_{b}}$ and $\rm{IV_{d}}$. The value for the threshold is derived from $\langle\chi_{\rm{red}}^2\rangle=\frac{\langle\Delta\chi^2\rangle}{N_{\rm{data}}-n-1}=1$, 
as $N_{\rm{data}}=78$ and $n=10$ correspond to the number of data points we fit and the degrees of freedom of our 
model respectively. Here, we emphasize that our criterion of chain abortion, especially the defined threshold 
$\Delta\chi^2_{\rm{abort}}$, is not universal. One should choose an appropriate solution depending on the $\chi^2$-distribution 
of the problem at hand and the required accuracy.

\begin{table}[ht]
 \centering
   \caption{Comparison of the best $\chi^{2}$ obtained with optimization starting with four different seeds 
            from six locations given in Table \ref{tab:startingpoint}.}
   \tabcolsep 5mm
  \begin{tabular}{ccccc}
     \toprule
     \toprule
chain & a      &  b     &  c      &  d    \\ 
      \hline
I     & 45     &  72    &  56     &  89   \\ 
II    & 82     &  39    &  12     &  68   \\ 
III   & 504    &  236   &  293    &  693  \\ 
IV    & 39     &  127   &  375    &  11  \\ 
V     & 34     &  262   &  165    &  31   \\ 
VI    & 85     &  52    &  127    &  64   \\ 
VII   & 106    &  31    &  527    &  22   \\
VIII  & 351    &  95    &  71     &  166  \\
   \toprule
   \end{tabular}
   \vspace*{2mm}
   \tablecomments{\textwidth}{(1) $\chi^{2}$ from SED-Fitting on the basis of database is 179.5.
   (2) The subscript letters from a to d represent different seeds used to initialize the pseudo-random 
       number sequence.}
\label{tab:x2result}
\end{table}

\subsection{Fitting Results}
Thirty-two CPUs were used to perform the optimization and each Markov chain was propagated on a single CPU.
The typical number of steps for a chain before abortion is ${\sim}1,200$. Because the calculation time 
per iteration highly depends on the optical depth of the proposed model, the total run-time of an 
individual chain can vary significantly. On average, it took about 25 days to complete one chain, 
comparable to the time to establish the database. The $\chi^2$, $\langle\Delta\chi^2\rangle$ and 
some model parameters of the Markov chains $\rm{II_{b}}$/$\rm{IV_{d}}$ are plotted in Fig. \ref{fig:tracechain} 
versus the accepted step count. It is clear from this plot that the merit function decreases quickly during 
the optimization run and then remains in the vicinity of ${\sim}50$. Both $R_{\rm{in}}$ 
and $M_{\rm{dust}}$ gradually converge to the value of the RM. This behavior is the base for the 
criterion described in Sect. 4.4 for chain abortion. Parameters like $R_{\rm{out}}$ that are difficult to be 
constrained by the observations, does not converge in an obvious pattern but instead cover an extended interval. 

The minimum $\chi^{2}$ for the best fit of each Markov chain are listed in Table \ref{tab:x2result}. 
By comparing these values to the minimum $\chi^{2}$ of ${\sim}179.5$ obtained from fitting the SED 
to the database we can evaluate the quality of fit. Most Markov chains found parameter sets with 
significantly lower $\chi^{2}$ as compared to the result from the database, demonstrating the 
applicability of SA in the field of circumstellar disk SED modeling.

More quantitatively, $75\%$ of the Markov chains found ``improved'' models compared to the best fit 
in our pre-computed database. Here, we use quotation marks to caution the reader that better $\chi^2$ not 
necessarily translates into disk parameters closer to the RM as the model is degenerate.

\begin{table}[ht]
 \centering
   \caption{The best-fit disk parameters for the RM found by SA and the derived error intervals.}
   \begin{tabular}{lc}
     \toprule
     \toprule
parameter  & best-fit   \\
   \hline  
\vspace*{2mm}
$T_{\star}$ [K]  &  $4010^{+38}_{-25}$    \\
\vspace*{2mm}
$L_{\star}$ [$L_{\odot}$] & $0.93^{+0.09}_{-0.06}$ \\
\vspace*{2mm}
$R_{\rm in}$ [AU] & $1.95^{+0.35}_{-0.14}$  \\ 
\vspace*{2mm}
$R_{\rm out}$ [AU] & $480^{+52}_{-28}$  \\
\vspace*{2mm}
$\alpha$ & $2.34^{+0.091}_{-0.141}$  \\
\vspace*{2mm}
$\beta$ & $1.24^{+0.069}_{-0.089}$  \\
\vspace*{2mm}
$h_{100}$ [AU] & $11.58^{+0.174}_{-1.715}$   \\
\vspace*{2mm}
$m_{\rm dust}$ [$\rm{10^{-5}M_{\odot}}$] & $4.57^{+1.701}_{-0.250}$  \\
\vspace*{2mm}
$a_{\rm max}$ [$\mu{\rm m}$] & $3.5^{+0.639}_{-0.134}$   \\
$i$ [$^{\circ}$] & $60.9^{+3.9}_{-2.2}$  \\
   \toprule	
   \end{tabular}
\label{tab:bestfitofsa}
\end{table}

\subsection{Estimation of {\it Local} Error Intervals with SA}
Estimation of the uncertainties for the best-fit parameters using SA are quite different from the previously described 
Bayesian analysis. First, a confidence interval $\chi^2_{\rm{conf}}$ has to be defined from already
calculated models. The deduction of this bound relies on the practitioners experience as no analytic solution can 
be given. Secondly, the vicinity of the best-fit in parameter space is probed by starting one or several Markov chains 
from this location. After collecting sufficient parameter sets below the confidence threshold, this normally 
asymmetric domain can be characterized by taking the minimum and maximum value of each degree of freedom. 

We started a new Markov chain from the best-fit model of chain $\rm{IV_{d}}$ to sample 
the $\chi^2$-distribution. The low starting temperature of $\tau{\sim}3$ (see Eq. 9) 
retained the random walk to the vicinity for sufficient steps to examine the local 
optimum. We gathered ${\sim}500$ samples with this new chain below the confidence 
threshold $\chi^2_{\rm{conf}}$ defined by 
\begin{equation}
\chi^2_{\rm{conf}}-\chi^2_{*}<3(N_{\rm{data}}-n-1),
\end{equation}
where $\chi^2_{*}$ is the new overall minimum $\chi^2$ found during the restart run (\citealt{robitaille2007}). 
Again, $N_{\rm{data}}=78$ and $n=10$ denotes the number of data points and dimensionality of the model.
Table \ref{tab:bestfitofsa} summarizes the best-fit model of chain $\rm{IV_{d}}$, the error bounds are 
deduced from the minimal and maximal components of all sampled parameter sets.

We like to emphasize that the error bounds deduced with this method have to be interpreted as a 
{\it local} confidence interval as it is only based on a Markov chain probing the direct 
environment of the best fit while the Bayesian analysis is more appropriate for a {\it global} 
error estimation by taking the whole database into account.

\section{Discussion}
There are generally two explanations for discrepancies between SEDs derived from the 
best-fit model and the RM. First, the SED is a spatially integrated observable and
high-dimensional disk models are often degenerate. Hence, fitting of such models
can only provide weak constraints on the disk structure (e.g., \citealt{robitaille2007}).
Experience from previous modeling campaigns unambigously 
indicates that additional observations, especially multi-wavelength,
high resolution images, are crucial to overcome model degeneracy (e.g., \citealt{pinte2008, sauter2009}).
Secondly, it is impractical to search on a sufficiently fine grid in parameter space as realistic
models usually contain too many dimensions. The SED fitting problem is therefore an excellent 
showcase to assess optimization methods for astrophysical modeling. We will discuss in the 
following paragraphs the highlights of our study on fitting with a pre-calculated database 
and simulated annealing.

\begin{figure}[!htp]
\centering
\includegraphics[width=\textwidth]{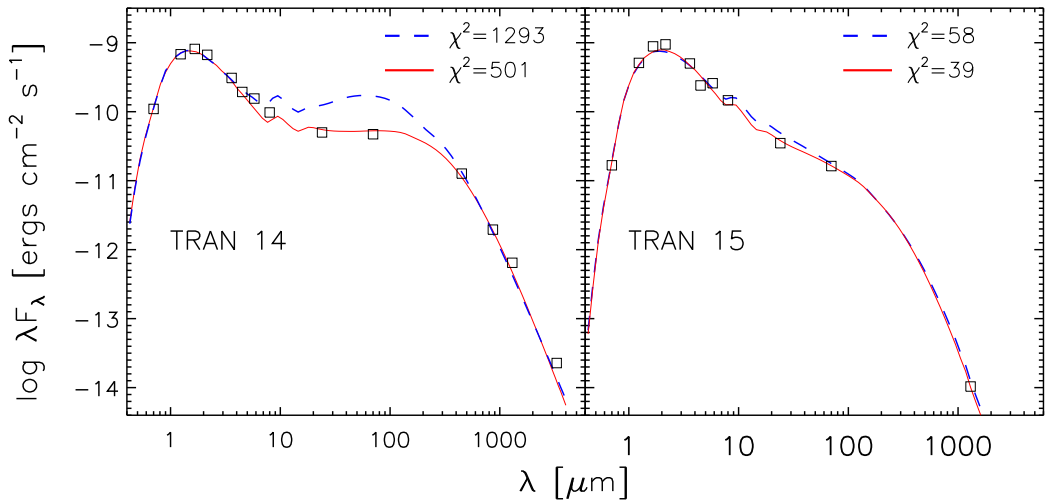}
\caption{An illustration of hybrid approach to model SEDs of two transition
              disks identified by \cite{cieza2010}, in their nomenclature TRAN 14 (left plot) 
              and TRAN 15 (right plot). The best-fit model in the pre-calculated database
              is indicated as a blue dashed line. The red solid line represents a highly improved
              model found by SA based on the results informed by the database. Observations
              are overplotted with black squares.}
\label{fig:hybridapproach}
\end{figure}

\vspace*{2mm}\noindent{\bf Database vs. SA:}\hspace*{3mm} We implemented two different approaches to
fit SED data of the RM, the use of a pre-calculated model database 
and SA, and compared these two methods extensively.

After construction of the database, a single set of observations can be fitted very fast by
calculating the $\chi^2$ of every model in the database. The result may be only an approximation
but the database enables us to deduce global error intervals for all parameters with the 
Bayesian inference method. It evaluates the probability for each parameter by weighing all models 
in the database. The fitting accuracy highly depends on the grid resolution of the models. 
Unfortunately, the number of grid points increases exponentially with the dimensionality of 
the parameter space. It is hence necessary to make a compromise between grid resolution 
and available computational resources.

Our results show that in most cases SA finds parameter sets with a $\chi^2$ approximately 
half the size of the solution from the pre-calculated database, indicating a higher fitting 
quality on average. However, some Markov chains are unable to reproduce 
the SED quite as well during the allocated timeframe as their siblings. This is caused by 
the stochastic nature of Monte Carlo methods and should be taken into account by using 
sufficient Markov chains in parallel to perform the optimization. The increasing 
computational speed of multicore CPUs ameliorates this embarrassing/parallel problem 
but the user will have to find a compromise to retain the calculation time in a tolerable range too.
SA is a sequential process and the Markov chain may converge slowly. The fitting of SED 
data for a sample of objects can therefore be a very time consuming task, even 
for a medium computer cluster with hundreds of CPUs.

Hence, using SA and a pre-calculated model database have their individual merits
for solving astrophysical fitting problems. SA is often better suited for individual cases 
whereas a database is preferable in a sample study, and the practitioner has to choose by 
weighing the computational constraints against the required accuracy. One can of course use a hybrid 
approach by taking the best-fit model from the database as starting point for SA. Moreover, the most 
probable value for each parameter indicated by the peak of Bayesian probability distribution
is also a good starting choice for SA. An illustration of this idea is given in Fig. \ref{fig:hybridapproach} 
for fitting SEDs of two ``transition disks'' in the Ophiuchus molecular cloud identified by \cite{cieza2010}, 
in their nomenclature TRAN 14 and TRAN 15. The quotation mark here is used to remind that the selection
criterion for transition disks used by \cite{cieza2010} is broad. Six different Markov chains starting 
from the best-fit, the most probable parameter set in our database or their surroundings were used to 
optimize the fit. After we performed ${\sim}150$ accepted steps for each Markov chain, the quality of the 
fit for both objects were highly improved.

\vspace*{2mm}\noindent{\bf Starting point for SA:}\hspace*{3mm}
The influence of the starting point on the probability to encounter the vicinity of the 
global optimum is obvious. In our study, the starting location III features the largest distance 
in parameter space to the RM, see Table \ref{tab:startingpoint}. Consequently, more steps 
are required to reach the RM. The corresponding chains therefore has more difficulty in 
finding better solutions than the best fit from the database compared with other optimization 
runs. As some parameters of astrophysical models can be constrained by observations, 
we suggest to choose the starting values accordingly.

\section{SUMMARY}
We have compared in detail the fitting of continuum SED using a pre-calculated database and SA, a Markov chain 
Monte Carlo method that has been successfully employed to solve many optimization problems outside the field of astrophysics. 
All the simulations are performed with the radiative transfer code \texttt{MC3D}.
Our study shows the practical application of SA in the context of circumstellar disk SED modeling and compares fitting 
quality and efficiency of both approaches. Because the same radiative transfer code, disk structure and dust composition 
are used to calculate the database and the Markov chain, the comparison of fitting 
quality reduces to a comparison of resulting $\chi^{2}$. We show that SA typically finds better 
solutions than the database, directly demonstrating the applicability of this algorithm to the modeling 
of SED data. However, a database can quickly evaluate the overall uncertainty of all parameters and 
provide a good starting point for SA. A hybrid approach to combine both methods leads to more 
accurate solutions.

\normalem
\begin{acknowledgements}
 Y.L. acknowledges support by the German Academic Exchange Service. 
 H.W. acknowledges the support by NSFC grants 10733030, 10921063, and 11173060.
 We acknowledge the anonymous referee for the comments and suggestions that helped to improve the paper.
\end{acknowledgements}

\bibliographystyle{raa}
\bibliography{bibtex}

\end{document}